\documentclass[%
 reprint,
 amsmath,amssymb,
 aps,
 amsfonts,
 mathrsfs
]{revtex4}

\usepackage{graphicx}
\usepackage{dcolumn}
\usepackage{bm}

\usepackage{algorithm}
\usepackage{algpseudocode}

\usepackage{xcolor}%
\usepackage{textcomp}%
\usepackage{booktabs}%
\usepackage{braket}

\begin{document}


\title{Global dynamical stability for KMS states}


\author{H. Narnhofer}
\affiliation{Institute of Theoretical Physics, University of Wien, Wien,  Austria}


\begin{abstract} The Hugenholz-Boltzmann evolution is generalized to strongly interacting systems on the lattice. Under appropriate assumptions states stable under this evolution are shown to satisfy the KMS-condition. How far these assumptions are reasonable is discussed.
\end{abstract}


\maketitle

\section{Introduction}

It is commonly accepted that equilibrium states of quantum spin systems satisfy the KMS-condition ~\cite{HHW},~\cite{BR}. However this has to be justified by showing that only these states satisfy demands that we expect to hold on physical grounds. It is our aim to reduce these demands as much as possible so that most of them turn out to be consequences of the underlying model and not demands.

A famous demand is passivity ~\cite{PW}. Based on the second law of thermodynamics for time invariant states perturbed locally both in space and time have gained energy, when the perturbation stops. We should notice that this demand is a consequence of observation and not the result of a thermodynamic limit, though the thermodynamic limit is necessary to show its equivalence to the KMS-property. Also sufficiently many perturbations have to be taken into account in the proof.

Another demand is dynamical stability ~\cite{HKTP}. Again perturbations of the dynamics are considered, now local in space but lasting for an infinite time. New states evolve. Based on the assumption of commutativity properties they tend to a limit and it is assumed that the limit is independent from the time direction. In the original paper strong multi-cluster-properties are assumed that however are not satisfied in general for interacting systems ~\cite{N2},~\cite{N3}. Fortunately it is possible to reduce these demands based on ideas in ~\cite{PW} and their relation to time reflection ~\cite{N4}.
Violation of dynamical stability indicates global effects in the cause of time.

As a minimal assumption to justify that equilibrium states are KMS-condition on the basis of a thermodynamic limit we use the fact that invariant means of initial states are time invariant. If we are able to show that for every initial condition the invariant mean or even better its time-limit satisfies the KMS-condition we are done. A result in this direction is given by the Boltzmann evolution in classical mechanics. Here the initial state is free of non-trivial correlations between the particles. The time-scale in which the particles respond to the interaction is negligible compared to the time scale in the Boltzmann equation and the effect of interaction can therefore be represented just by the scattering matrix between two particles that enters in the Boltzmann equation. Also the particle density is so small that simultaneous interaction between more than two particles can be ignored. Under these conditions the system evolves to a state that is characterized by a temperature. So far the Boltzmann equation is an ansatz for a time evolution. That it follows in the limit of increasing particle numbers is shown under various appropriate assumptions on the initial state at least for a limited time ~\cite{Pu}, ~\cite{OL}, ~\cite{IP}.

In ~\cite{Hu} an analogue to the Boltzmann equation for quantum lattices is constructed. The initial state is quasifree or sufficiently regular to converge under a quasifree evolution to a quasifree state. The quasifree dynamics is perturbed by an interaction that is sufficiently weak so that in finite time the system does not react on it in an observable manner. But the strength of the interaction is adjusted to the time that the system needs to converge to its time average so that in this appropriate time scale the interaction becomes effective. In this scaled limit similarily as in the Boltzmann equation the state tends to a limit and this limit-state satisfies the KMS condition. More precisely the Hugenholz equation becomes a non linear equation between the space-averaged two-point functions and this is sufficiently explicit to cheque the KMS-condition.

In this paper we adapt the approach of ~\cite{Hu} to strongly interacting systems. Using the cluster-properties that are essential for the validity of the Hugenholz-dynamics we assume that they also hold for time-invariant states of interacting systems. Now we perturb the time-evolution e.g.
by a quasifree evolution with a scaling mimicking that of ~\cite{Hu}, where again the perturbation is space translation invariant. Under the appropriate assumptions the time-invariant state reacts on the perturbation only up to second order terms. If we assume that also up to this order the perturbation vanishes, as it happens in the limit state of ~\cite{Hu}, then we show that the state is in fact a KMS-state.

Differently to quasifree evolution we do not know in general which other time-invariant states exist and therefore we do not obtain an explicit evolution of such states in an appropriate scaling. For interacting systems we can construct time evolutions that commute with the given one and therefore can use their KMS-states as states also invariant for the given time-evolution. On these states the resulting Hugenholz-equation can be applied. However for all invariant states that are not KMS-states with respect to the time-evolution under consideration  they are not stable under minimal perturbations of the dynamics , especially under minimal variations of the coupling constant. Only KMS-states remain stable under this minimal variation that however does not suffice to give a variation of the dynamics in finite time on the level of C*-algebras.

\section{The Hugenholz-dynamics and the necessary correlation inequalities}
We repeat the approach of ~\cite{Hu} as far as we will generalize it for interacting systems. Consider a Fermi system on an infinite lattice with a time evolution given by a hamiltonian $H=K+\lambda V$ where $K$ is quadratic in creation and annihilation operators and corresponds to the kinetic part and $V$ is of forth order in creation and annihilation operators and determines the interaction. Both are gauge invariant. The time evolution is constructed as in ~\cite{BR} and commutes with space translation. Without interaction the time invariant states are quasifree and are defined by a space-translation invariant two-point-function. We start with such a state $\omega _0$ and let it involve under the time evolution to $\omega _t ^{\lambda}$ which can be calculated on the level of perturbation theory.
Consider \begin{equation} \omega _{\tau}=\lim _{t\rightarrow \infty, \lambda ^2 t=\tau }\omega _t^{\lambda }.\end{equation}
On the basis of perturbation theory it is shown that this limit exists , that $\omega _{\tau }$ is a space translation invariant quasifree state, therefore also invariant under the time evolution without interaction. For a space translation invariant state the two point function corresponds to a $\rho (p)$ given by \begin{equation} \omega (a^*(f)a(g))= \int dp f(p)\bar{g}(p) \rho (p) .\end{equation}
With respect to $\tau $ it corresponds to a differential equation for this function $\rho (p)$ quadratic in $\rho (p)$. This differential equation has a stable solution $\rho _0(p)$ that satisfies a KMS-condition with respect to $K$.
In the expansion of $\omega _t^{\lambda }(A)$ the first order term in $\lambda $ vanishes for $t\rightarrow \infty$ in the relevant scaling because of the time clustering and the fact that the potential locally vanishes. The second term remains in the scaling. The convergence is guaranteed under sufficiently strong cluster properties.

For strongly interacting systems we start with a hamiltonian
\begin{equation}H_{\lambda }=K+V+\lambda H'.\end{equation}
Here $H'$ is a space translation invariant operator corresponding to a short range operator $H_0'$, that per perturbation theory implements an automorphism It must not commute with $H_0,$ otherwise we have some freedom and can choose it e.g. quasifree. This hamiltonian implements an automorphism $\alpha _{\lambda }(t)$ on the algebra. We start in analogy with ~\cite{Hu} and a state for which we assume that
\begin{equation} \lim _{\rightarrow \infty } \omega (\alpha _0(t)A)=\omega _0(A). \end{equation}
exists and is invariant under $\alpha _0$ as well as under space translation. Further we assume that
\begin{equation} \omega _{\tau } (A)= \lim _{t\rightarrow \infty , \lambda ^2 t=\tau}\omega (\alpha _0(-t)\alpha _{\lambda }(t)A)\end{equation}
exists and is independent with respect to $\alpha _0(t')$ for finite $t'$. Especially we demand that the state satisfies multiclustering in $t$:
\begin{equation}\lim _{T\rightarrow \infty , t_j -t_{j-1}>T} \omega (\alpha _0(t_1)A_1 \alpha _0(t_2)A_2..\alpha _0(t_n)A_n)=\omega (A_1)..\omega (A_n) \end{equation}

The limit state satisfies the differential equation
\begin{equation}\frac{d}{d\tau }\omega _{\tau } (1/2\int dt_1 \int dt_2 [H_{t_1}'H_{t_2}'A+A H_{-t_1}'H_{-t_2}'-2H_{t_1}'AH_{-t_2}'])\end{equation}
with $H_t'=\alpha _0(t)H'.$ It is based on perturbation theory  where we observe that the first term vanishes in the chosen limit, due to the clustering in $t$, the second term contributes in areas of finite $t_1,t_2$ and is integrable. In expanding in $\lambda $the terms of order $\lambda^n$ are sums over
\begin{equation} \Pi _{k, t_l }\alpha _0 (t_l)\sigma _{j+k}H_j', \quad 0<k<n \end{equation}
with $j$ running over the whole space and the obvious conditions on $t_l$ corresponding to the expansion of the unitary. As operator these terms are not summable. But we demand  multi-clustering as property of the state in the sense that most of the terms factorize into their expectation value in the state and only those that after being shifted both in space and time remain being close to the local operator contribute according to the appropriate summability. The summation over the terms that factorize cancel in their action on the local operator whose evolution we study. Therefore only terms of order $\lambda ^{2n}$ remain, when we take the scaling into account. Then the expansion is summable and is the solution of the Boltzmann-Hugenholz- differential equation.

We are interested in characterizing a state that is invariant under this evolution. With $W=\int _0 ^  {\infty } dt H'_t$ this state satisfies in the GNS-representation
\begin{equation}\langle \Omega | WWA+AW^*W^*-2WAW^*|\Omega \rangle =\langle \Omega |A(W^*W^*+\Delta WW+W^*\Delta W)|\Omega \rangle =0\end{equation}
with $\Delta $ the modular operator corresponding to $\omega.$ Since $A^*|\Omega \rangle $ is dense in the Hilbertspace it follows that
\begin{equation} (W^*W^*+\Delta WW+W^*\Delta W|\Omega \rangle =0.\end{equation}
$H$ implementing the time evolution commutes with $\Delta $ in a time-invariant state.  Therefore we can express the operator in the joint spectral representation $\mu, \lambda $ and can ignore additional refinement of the spectrum since $H' $ is selfadjoint and obtain for the relevant vector that has to vanish
\begin{equation} \int d\lambda ' d\mu'[(1+e^{-\lambda })W(\lambda,\lambda'; \mu, \mu ')W \lambda ',0;\mu',0)\frac{1}{(\mu -\mu '-i\epsilon )(\mu '-i\epsilon)} \end{equation} $$-2W(\lambda ,\lambda ';\mu, \mu ')e^{-\lambda '}W(\lambda ',0; \mu ',0)\frac{1}{(\mu -\mu '-i\epsilon ) (\mu '+i\epsilon)}]$$
The vector vanishes only if the singularities in the integral cancel one another.  This happens only if the integral contains an additional $\delta (\lambda -\beta \mu)$, so that the singularity at $\mu' =0$ gives $\lambda '=0$ and cancels $1$ and the singularity at $\mu '=\mu $ cancels $e^{-\mu }=e^{-\lambda }.$  Thus the state has to be a KMS-state for some temperature $\beta ^{-1}.$

It follows that again the Boltzmann-Hugenholz-evolution also for this generalized application for strongly interacting systems only allows KMS-states as states invariant for sufficiently long time under sufficiently small though not vanishing perturbation, where only one type of perturbation has to be taken into account.

\section{The Boltzmann-Hugenholz-evolution as a Lindblad evolution}
The Boltzmann- Hugenholz-equation is an evolution of Lindblad-type ~\cite{L}, ~\cite{AL}, i.e. it defines a semigroup on the operators of completely positive type that preserves states, so that we only generalize the condition, that density matrices are mapped into density matrices. In ~\cite{BCFN} such evolutions have been studied for mean field evolutions. There the operator $W$ is scaled as $N^{-1/2}$ in the thermodynamic limit, otherwise translation invariant, so that $N$ corresponds to the scaling of $t$ for given $\tau.$ For such a mean-field evolution $[W,W^*]$ converges to a c-number. This corresponds to the demand that in our approach $\omega ([W,W^*])$ converges, following from appropriate time clustering.

An important observation both for the Boltzmann-equation as for the Hugenholtz-equation for quasifree dynamics is the fact that the entropy increases. This can be observed, because the entropy density is defined by the two-point function. We do not know $\omega _{\tau }$  sufficiently explicit to make statements about its entropy density. For Lindblad evolutions entropy can both increase and decrease. General statements are only available if the Lindblad equation acts on density matrices where we can control the evolution of the eigenvalues.
Every state corresponds to a density matrix on the local level $\rho _X.$ Let us replace $W$ by $W_X$ where $W_X$ is also localized in the region $X.$ We consider
\begin{equation} \frac{d}{d\tau}\rho _X (\tau )=- W_X W_X\rho _X(\tau )-\rho _X(\tau )W^*_X W^*_X +2W^*_X\rho _X (\tau )W_X \end{equation}
Let the eigenvalues of $\rho _X (\tau )$ be $r_j$ with corresponding eigenvalues $|j\rangle .$ Then its entropy is given by $-\sum _j r_j \log r_j.$ Write

\begin{equation} W=w_{jk}|j\rangle \langle k|\end{equation}
Then
\begin{equation} \frac{d}{d\tau }S(\rho _X (\tau )= \sum _{jk} (|w_{kj}|^2 r_j \log r_j -|w_{jk}|^2 r_j\log r_k)\end{equation}

With $r_j \log r_j \geq r_j-r_k +r_j \log r_k$ and $\sum _j r_j=1$ it follows that the entropy increases.
When we replace $\omega $ by $\bar{\omega }=\omega _{X^c}\otimes \omega _X $ then we can take the conditional expectation of $W$ with respect to $\bar{\omega }$ and obtain $W_{X^c}\otimes W_X$ which differs from $W$ only by a surface term with respect to the region $X.$ Calculating the entropy of the region $X$  we either have to take into account the restriction of $\omega _{\tau }$ to the region $X$ or the change of the restricted state by the Lindblad-evolution created by $W_X.$ The two states differ by a surface term and therefore the difference of the entropies is of surface-size, whereas the change of the entropy by the localized Lindblad-evolution is of the size of the region. In the thermodynamic limit the increase is dominating and the entropy-density of $\omega _{\tau }$ is increasing under the Boltzmann-Hugenholz- evolution.

\section{Are the assumptions plausible?}
In \cite{Hu} all considerations are based on our knowledge of quasifree time evolution and the corresponding evolution of quasifree states. Especially we can start with a space translation invariant state that is also time translation invariant and all its time correlations are under control by the spectral properties of the time evolution in the two-point function.

We can replace this approach if we know whether other time invariant states exist and if we can make statements on their correlation functions.
For a time invariant state that is faithful the modular time evolution exists and defines an automorphism for the corresponding von Neumann- algebra and for this algebra time automorphism and modular evolution commute \cite{N}. If we can construct another time evolution on the $C^*$-algebra that commutes with the given time-evolution then we can use its KMS-state as starting point for the Boltzmann-Hugenholz-evolution. The automorphisms commute if their derivatives commute and it is sufficient to cheque this commutativity on the action on just one creation operator. Depending on the chosen candidate for the commuting automorphism this asks for an increasing number of controls of polynomials in creation and annihilation operators.

Example: Let the time evolution $(\alpha (t)$ correspond to the derivatives $\delta _K, \delta _V$ with
\begin{equation} \delta _K a_0=[K_0,a_0], \quad \delta _V a_0=[V_0,a_0]\end{equation}
Let $\bar{\alpha }(t) $ correspond to the derivative \begin{equation}\gamma \delta _K +\frac{1}{\gamma } \delta _V.\end{equation}
Then
\begin{equation} \delta \bar{\delta }-\bar{\delta } \delta =0\end{equation}
The question arises, if this example can be generalized. E.g. we can choose $\delta _{\bar{K}} $ to commute with $\delta _K$, that holds if both derivations are quasifree and $\delta _V$ to commute with $\delta _{\bar{V}}$, which is satisfied if $V=\sum _k\sum _j\Pi _j V(j)\sigma _k\sigma _ {k+j}$ and similarly with $\bar{V}$ so that also the derivation with respect to the interaction commute. It remains to control whether
$$ \delta _K \delta _{\bar{V}} = \delta _V \delta _{\bar{K}}.$$
Comparing the polynomials that are involved it turns out that again only the choice (16) satisfies the demands.
For more general interactions already in $\delta _V \delta _{\bar{V}}$ in general polynomials appear for which there is no counterpart from other contributions . Therefore the offered example seems to be quite general, but is possible for every time evolution. Even more it allows to split $H=H_1+H_2$ and consider as commuting automorphism $\lambda H_1+\lambda ^{-1}H_2$ to construct other time-invariant states.
Comparing the polynomials that are involved it turns out that again only the choice (16) satisfies the demands.

Now we turn to the question whether the demand on the cluster properties can be satisfied. Instead of starting from an arbitrary state that converges to a time invariant state we start from the very beginning with a KMS-state with respect to an automorphism that commutes with the time automorphism. For such a KMS states some results on clustering properties are available. We refer to ~\cite{M}. We restrict ourselves to algebras on one dimensional lattices. Then there exist constants $K,M$ such that
\begin{equation} |\omega (Q_1 \sigma _jQ_2)-\omega Q_1)\omega (Q_2)|<K e^{-Mj}||Q_1 || ||Q_2|| \end{equation}

for $Q_1 $ in $\mathcal{A} _{[-\infty,0]}$, $Q_2$ in $\mathcal{A}_{[1,\infty ]}$. This can be extended to
\begin{equation} |\omega (Q_1 \sigma _j(Q_2 \sigma_j Q_3(..\sigma _jQ_n )- \omega (Q_1)..\omega (Q_n)|<nKe^{-Mj}, \quad Q_l \in \mathcal{A}_[a,b] \end{equation}

So far we can rely on multiclustering with respect to space-translation. But we need multiclustering with respect to both space translations and time translation. We refer to the result of \cite{LR} that interacting systems have finite group velocity
\begin{equation} ||[\sigma _x\alpha _t A,B]<e^{-|t|(\frac{\mu |x|}{|t|}-c}||A|| ||B||\end{equation}
where $A,B$ are localized in $\mathcal{A}_0$ and $\mu ,c$ are fixed by the time evolution, so that $\mu $ gives a finite velocity with which information can be spread. This inequality covers also the case when the unperturbed time evolution is periodic, therefore its group velocity is $0$. In this case we can take the average in time over this period. We can replace the differential equation by considering these finite steps. Then the small perturbation commutes with the unperturbed time evolution. Starting with a state that is invariant under the unperturbed evolution but with nontrivial spacial correlations the state evolves under the perturbation in a time scale of order $\lambda ^{-1}$, therefore in a time scale that is smaller than the one in the Boltzmann-Hugenholz evolution. The state changes faster, but the evolution is determined only by the perturbation, e.g.  with the appropriate choice of the perturbation corresponds to a quasifree evolution and the state will converge to a quasifree evolution but in general not to the KMS-state of the unperturbed evolution with its trivial spacial correlations. This shows that the results of ~\cite{LR} are promising but not sufficient to guarantee convergence to equilibrium under reasonable initial conditions.

We turn to our model of an interacting system in combination with a quasifree evolution. It can be shown \cite{N2},\cite{N3}, that the evolution is not periodic but the hamiltonian in an invariant state has continuous spectrum. This is not sufficient to guarantee weak convergence to an invariant state, better control on the dynamics would be necessary. However in \cite{CR} in the restricted situation of the vacuum state with quite generality on the interaction a lower bound on the signal propagation is given, expressed as a control on clustering in combination of space translation and time translation. If in addition we take into account the result in \cite{BBN} where a discrete automorphism group with finite speed but chaotic delocalization only admits the tracial state as invariant state that satisfies the desired multiclustering, we believe that there are good chances that for concrete examples of interacting systems the desired multiclustering in invariant states can hold or can even be proven on the basis of the Ruelle-Transfer-method.

\section{Conclusion}
Based on the assumption that clustering properties with respect to space and time are similar for time evolution with or without interaction
it is shown that global perturbation of the time evolution has an effect for the state that is only observable in a time scale larger than the inverse of its strength of the perturbation. In the appropriate scaling its effect corresponds to a Lindblad evolution for time-invariant states. For this Lindblad evolution the entropy-density is increasing. Limit states satisfy the KMS-condition. Therefore KMS-states are the only states that are stable under global perturbation of the dynamics  under a time scale that tends to infinity with respect to the time scale of the perturbation.


\begin{thebibliography}{999}
\bibitem{HHW} R. Haag, N.M. Hugenholz, M. Winnink: Commun. Math. Phys. {\bf5} 215 (1967)
\bibitem{BR} O. Bratteli, D.W. Robinson: Operator Algebras and Quantum Statistical Mechanics I and II, Springer (1981)
\bibitem{PW} W. Pusz, S.L. Woronowicz: Commun. Math. Phys. {\bf58} 273 (1978)
\bibitem{HKTP} R. Haag, D. Kastler, E. Trych-Pohlmeyer: Commun. Math. Phys. {\bf56}214 (1977)
\bibitem{N2} H. Narnhofer: Interacting systems in the tracial state, arXiv 2406.04977
\bibitem{N3} H. Narnhofer: Asymptotics for quantum systems on the lattice, arXiv 2405.19696
\bibitem{N4} H. Narnhofer:Journ. Phys. A:Math. Theor. {\bf41} 335211 (2008)
\bibitem{Pu}M. Pulvirenti: Boltzmann equation (Classical and Quantum) Encyclopedia of mathematical Physics 306-312 (2006)
\bibitem{OL} O. Lanford III: in Springer Lecture Notes {\bf35} (1975) 1-111
\bibitem{IP} R. Illner, M. Pulvirenti: Commun. Math. Phys. {\bf121} 134 (1989)
\bibitem{Hu} N.M. Hugenholz: Journal of Statistical Physics {\bf32/2} 231 (1983)
\bibitem{L} G. Lindblad: Commun. Math. Phys. {\bf48} 119 (1976)
\bibitem{AL} R. Alicki, K. Lendi: Quantumm Dynamical Semigroups and Applications. Lect. Notes Phys.{\bf717}Springer-Verlag Berlin (2007)
\bibitem{BCFN} F. Benatti, F. Carollo, R. Floreanini, H. Narnhofer: J. Phys.A: Math. Theor. {\bf50} (2017) 423001
\bibitem{N} H. Narnhofer: Acta Phys. Austr. {\bf47} (1977) 1
\bibitem{M} T.Matsui: Ann. Henri Poincare {\bf4} (2003) 63-83
\bibitem{LR} E.H. Lieb, D.W. Robinson: Commun. Math. Phys.{\bf28} (1972) 251-257
\bibitem{A} H. Araki: Publ.RIMS, Kyoto Univ {\bf20} 277 (1984)
\bibitem{CR} C. Radin: Commun. Math. Phys. {\bf62} 159 (1978)
\bibitem{BBN}B. Baumgartner, F. Benatti, H. Narnhofer: Journ. Phys.A: Math. Theor. {\bf43} 115301 (2010)



\end{thebibliography}
\end{document}